\documentclass{biophys-new}
\usepackage[utf8]{inputenc}
\usepackage{graphicx}
\usepackage[colorlinks,allcolors=cyan!70!black]{hyperref}
\usepackage[separate-uncertainty=true]{siunitx}
\UseRawInputEncoding
\DeclareSIUnit{\litre}{l} 
\DeclareSIUnit\molar{\mole\per\liter}
\DeclareSIUnit\Molar{\textsc{m}}
\usepackage{lineno}
\usepackage{multirow}
\usepackage{lastpage}
\usepackage{fancyhdr}
\usepackage{ulem}
\bibstyle{natbib}
\usepackage{lipsum}
\title{\LARGE{Women and the Nobel Prize in the sciences: an update to 2026}}
\runningtitle{Preview} 
\author[]{Per Lunnemann}
\author[]{Liselotte Jauffred}
\runningauthor{Lunnemann and Jauffred} 
\affil[]{jauffred@nbi.dk}

\papertype{Commentary}
\begin{document}
\begin{frontmatter}
\vspace{0.35cm}

\begin{abstract}
Every October, the Nobel Prizes in Physics, Chemistry, Physiology or Medicine and Economic Sciences are announced, with one to three prizes in each discipline. Of the 769 Laureates (since 1901) in these four disciplines only 4\% are women (30 Laureates). In 2026, there are eight Laureates within these disciplines and they are all men. Based on historical data across four fields and a Bayesian hierarchical model, we update the 8 year old model on bias in Nobel Prizes \cite{Lunnemann2019GenderPrizes} and confirm the gender gap, with exceedingly large confidence.
\end{abstract}
\end{frontmatter}
\begin{figure}[b]
    \centering
    \includegraphics[width=1\linewidth]{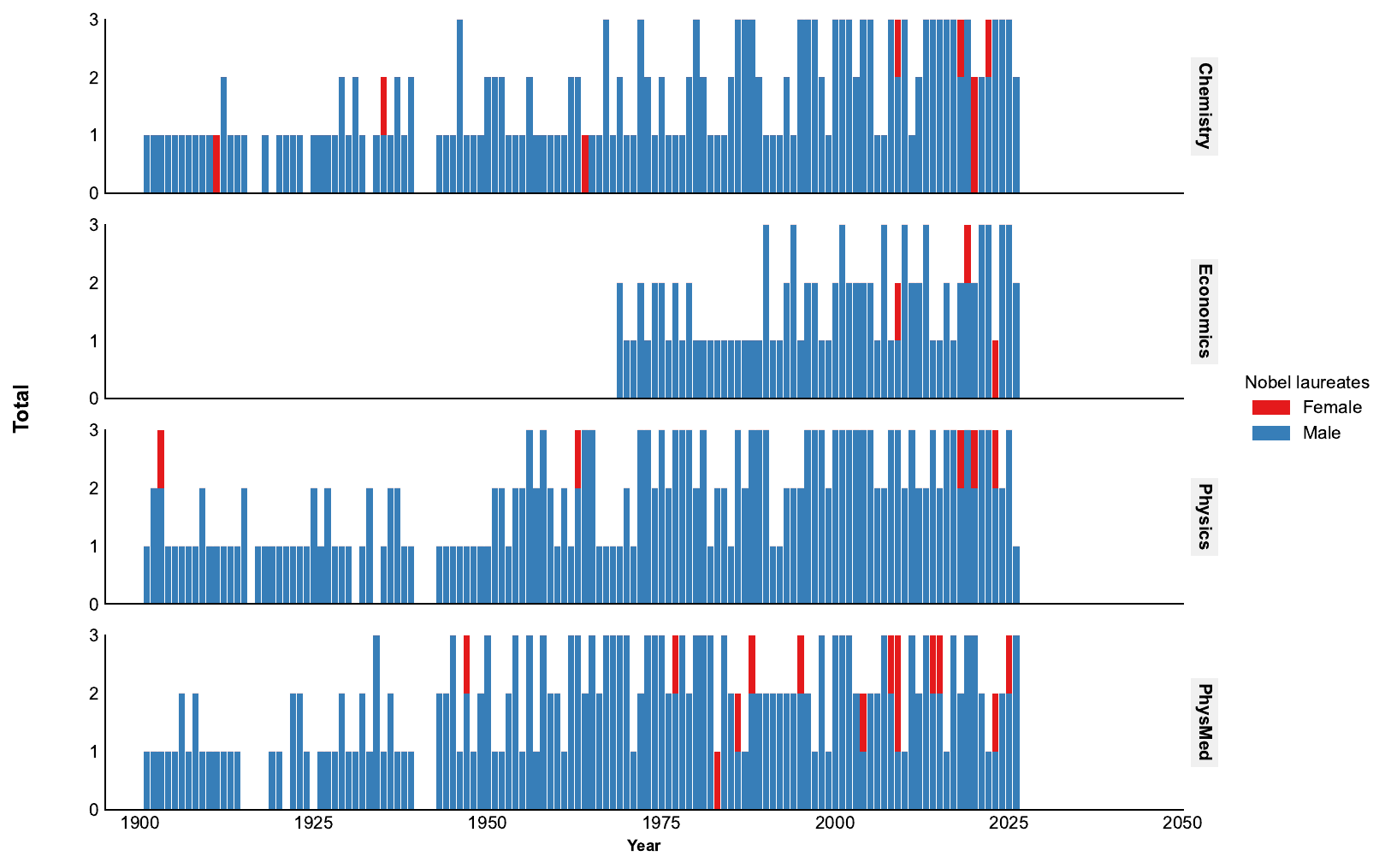}
    \caption{Gender distribution of Nobel Prizes. Bar plot of the scientific Nobel Prizes from 1901 to 2026 resolved by field and gender ratio, $r$.}
    \label{fig:NP}
\end{figure}
\begin{figure}
    \centering
    \includegraphics[width=0.8\linewidth]{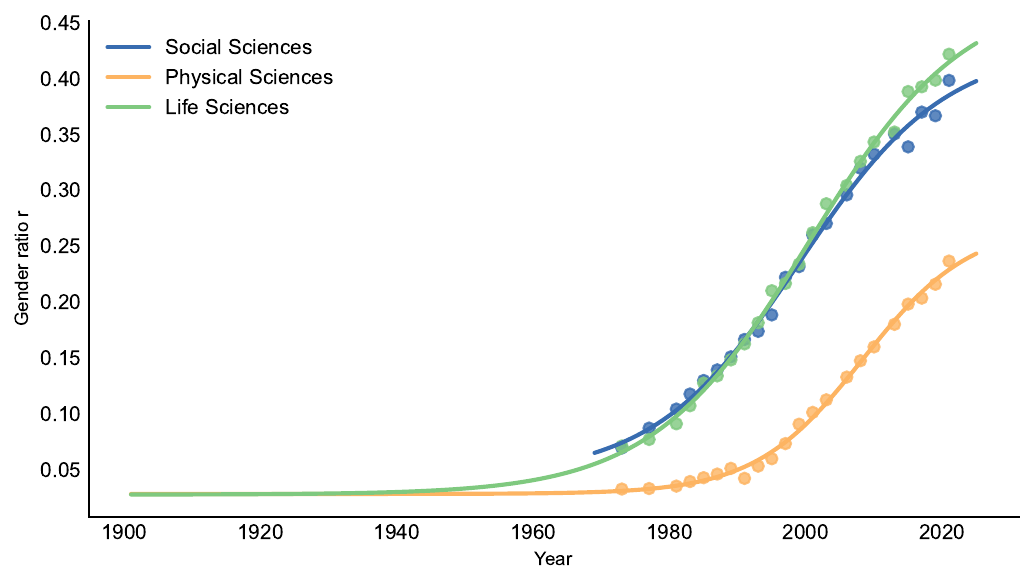}
    \caption{Gender ratio, $r$, of senior faculty versus year. $r$ is the number of women relative to all senior faculty members. Points are data from the National Science Foundation (Ref \cite{NSF2018}; 1973-2021). Lines are logistic fits, extrapolated back to 1901 for Physical and Life sciences and to 1969 for Social sciences, which is the first year of the prize in Economics. Physics and Chemistry share the Physical sciences ratio, and Economics uses the Social sciences ratio.}
    \label{fig:r}
\end{figure}

%
Women have historically held far fewer positions in academia than men, so fewer women Laureates are expected, but does this fully account for the gap? In 2019, we found that it did not: in all four disciplines, women were awarded the prize less often than the gender ratio suggests \cite{Lunnemann2019GenderPrizes}. 
The share of women among senior faculty has continued to rise (figure \ref{fig:r}) and so has the frequency of women Laureates (figure \ref{fig:NP}). So, here we ask whether eight more years change the conclusion?

The relevant reference gender ratio, $r$, is that of the pool the Laureates are drawn from. We therefore use the share of women among senior faculty as a proxy for the relevant group of researchers. Because the prize typically rewards work done decades earlier, and because the odds of a prize increase with age, we use the gender ratio 15 years before each award \cite{Bjork2019}. Hence, if a field had 10\% women 15 years before, we would expect a 10\% chance that any given Laureate is a woman, and the number of women Laureates would follow a binomial distribution. Since the number of Laureates per year is small ($\leq 12$), no conclusion can be drawn from a single year. We therefore ask how likely the observed number of women is, given the pool, over the full period.
\\ \\
We treat each \textbf{Laureate as a random draw} from the senior faculty of the field some years earlier, so that a priori sets the probability of a woman equals the share of women in that cohort. Introducing a bias parameter, $\alpha$, allows for scaling of this probability. If $\alpha$ is less than (more than) one, women are awarded less (more) often than their ratio suggests. As a metric, we report the probability that $\alpha$ is at least one (i.e., women are not disfavoured).
Similarly to the original analysis, Physics and Chemistry share the gender ratio of Physical sciences, and Economics uses that of Social sciences.

\begin{figure}
    \centering
    \includegraphics[width=0.8\linewidth]{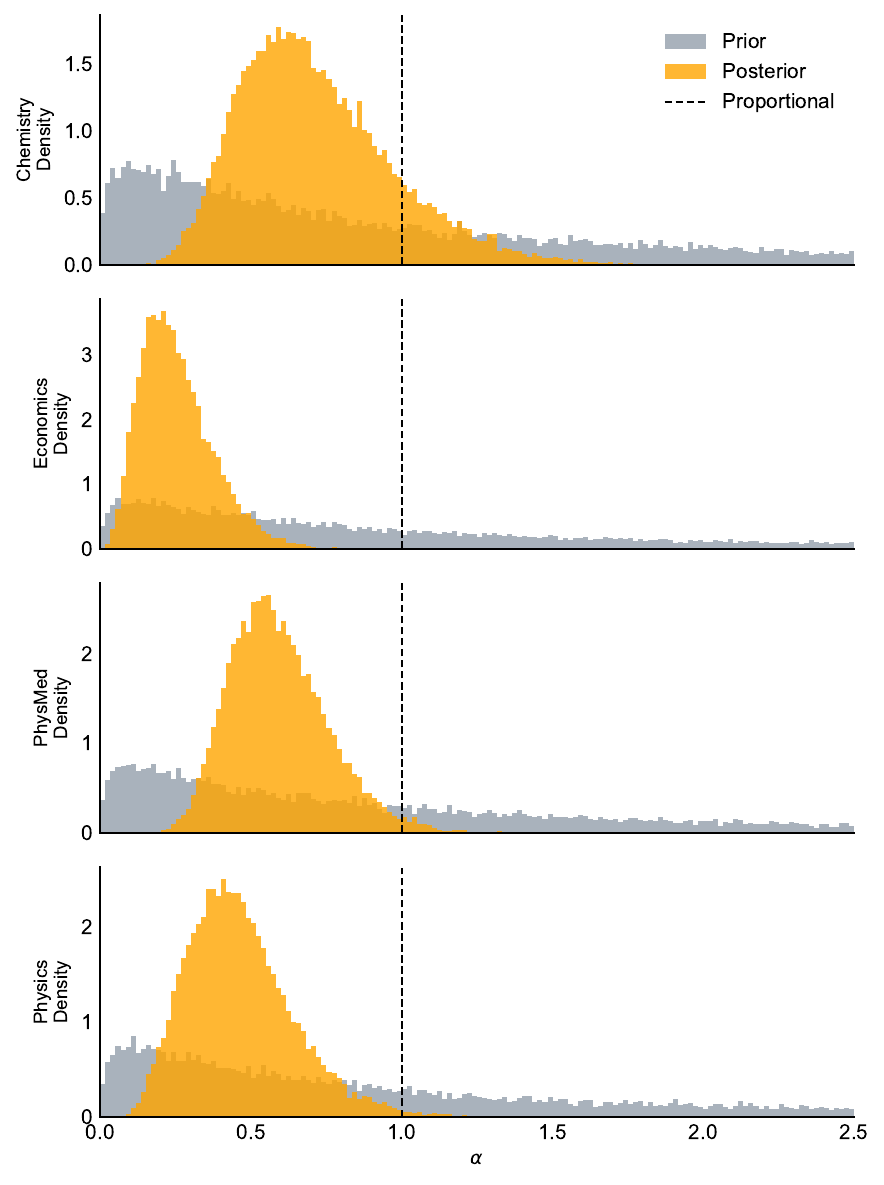}
    \caption{Prior (grey), posterior (orange) and proportional probability densities of $\alpha$, for $\delta = 15$. The prior has a median of 1, see Eqs. (3) and (4) in Ref \cite{Lunnemann2019GenderPrizes}. $\alpha < 1 (\alpha > 1)$ means women receive fewer (more) Nobel Prizes than the gender ratio suggests.}
    \label{fig:yellow}
\end{figure}

\begin{figure}
    \centering
    \includegraphics[width=0.8\linewidth]{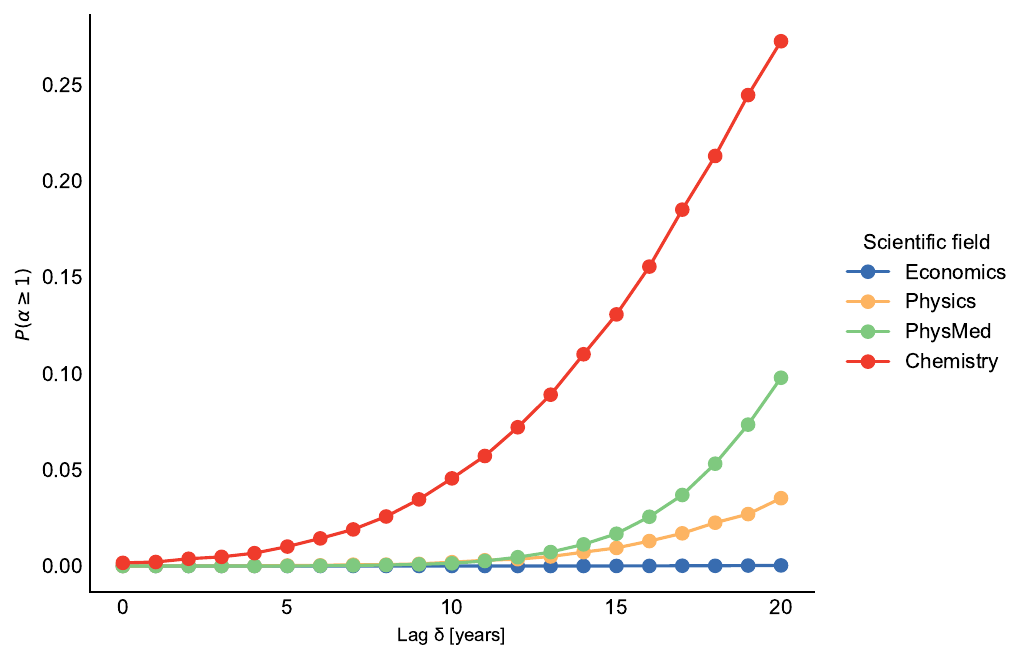}
    \caption{Probability that $\alpha$ is at least one, $P(\alpha\geq 1)$, versus lag, $\delta$. $P(α ≥ 1) = 1 − ∫₀¹ p(α|r, δ)d\alpha$ is the probability that women are not disfavoured.}
    \label{fig:bias}
\end{figure}

For all four disciplines, the posterior of the bias parameter $\alpha$ lies predominantly below unity (figure \ref{fig:yellow} for $\delta = 15$, and figure \ref{fig:bias} for $0 \leq \delta \leq 20$). That is, still to this date, women receive fewer prizes than their gender ratio suggests. Figure \ref{fig:bias} shows the probability that women are not disfavoured, $P(\alpha \geq 1)$, versus the lag $\delta$. For Economics, $P(\alpha \geq 1)$ is close to zero for all lags. For Physics, it stays below 4\% for all lags up to 20 years. For Physiology or Medicine, it is below 2\% for $\delta \leq 15$ and rises to $\approx 10\%$ at $\delta = 20$. For Chemistry, it is $\approx 5\%$ at $\delta = 10$, $\approx 13\%$ at $\delta = 15$ and $\approx 27\%$ at $\delta = 20$.
\\ \\
Thus, the \textbf{conclusion of the original analysis holds} for Economics, Physics and Physiology or Medicine. For Chemistry, the evidence is weaker than in 2019. The three women awarded in Chemistry since our original analysis (Charpentier and Doudna, 2020; Bertozzi, 2022) account for part of this change. Moreover, we compare Chemistry with the gender ratio of Physical sciences, which includes Physics, where the share of women is lower. Hence, we likely underestimate the bias in Chemistry. We therefore do not conclude that the bias in Chemistry has disappeared, only that the data no longer exclude it as firmly.
\\ \\
The natural \textbf{next question} is where in the process the gap arises. Nominators and nominees are not disclosed before 50 years after a price has been given, and therefore little is known about the selection. A study in Science \cite{KatieLangin2021OneNominees} gives a partial picture, with 13\% women among nominees in Physiology or Medicine and 8\% in Chemistry. If these are the pools from which the Laureates are drawn, the outcome is not surprising. With 8\% per Laureate, the probability that two women share a prize, as in Chemistry in 2020, is below 1\%, and a year without a woman is the expected outcome. The nomination data therefore give little reason to believe that the committees avoid women. It is more likely that the selection takes place before the nomination.
The prize is a delayed record of a pipeline. The share of women has increased, from 15 women in the first 100 years of the prize to 15 in the last 25, but the prize reflects who was in the cohort 10-15 years earlier. Perhaps a more informative question is thus not who the committees choose, but who gets to be in the run-up, and what it requires for women to remain there.
\clearpage
\section*{Model}
We use the share of women among senior faculty in the US, resolved by field, from the National Science Foundation \cite{NSF2018} as a proxy for a global distribution. The data cover 1973--2021. To obtain the gender ratio, $r$, for the full period of the prizes, we fit a logistic function to each field and extrapolate back to 1901 (1969 for Social sciences, the first year of the prize in Economics). The data group Physics and Chemistry under Physical sciences; hence, we use this ratio for both. For Economics, we use the ratio of Social sciences.

Laureates are typically senior researchers, and the awarded work was, on average, done 15 years before the prize. We do not have the gender distribution resolved by age. We therefore introduce a lag time, $\delta$, and assume that the Laureates in year $i$ are drawn from senior faculty in year $i-\delta$.

We model the number of women Laureates, $f_{ij}$, in field $j$ and year $i$ as a binomial variable:
\begin{equation}
f_{ij} \sim \mathcal{B}(N_{ij}, \theta_{ij}),
\end{equation}
where $N_{ij}$ is the number of Laureates and $\theta_{ij}$ is the probability that a Laureate is a woman. In the case of no bias, $\theta_{ij} = r_{i-\delta,j}$. To quantify any bias, we write
\begin{equation}
\theta_{ij} = \mathrm{logit}^{-1}\!\left[\mathrm{logit}(r_{i-\delta,j}) + \log(\alpha_j)\right],
\end{equation}
where $\alpha_j > 0$ is a time-independent bias parameter for field $j$. For $r \ll 1$ and $\alpha_j r \ll 1$, this reduces to $\theta_{ij} \approx \alpha_j r_{i-\delta,j}$. Hence, $\alpha_j = 1$ means that women are awarded the prize exactly as often as the gender ratio suggests, while $\alpha_j < 1$ ($\alpha_j > 1$) means less (more) often.

We assume some similarity between the four fields and use a hierarchical structure, where $\log(\alpha_j)$ is drawn from a normal distribution with hyper-parameters $\mu$ and $\sigma$:
\begin{align}
\log(\alpha_j) &\sim \mathcal{N}(\mu, \sigma), \\
\mu &\sim \mathcal{N}(0, 1), \\
\sigma &\sim \mathcal{N}(1, 0.5).
\end{align}
For $\mu = 0$, the median of $\alpha_j$ is 1, corresponding to no bias. The prior is thus weakly informative and allows values of $\alpha$ well below and above 1 (figure \ref{fig:yellow}). [The results are robust to the choice of the hyper-priors on $\mu$ and $\sigma$.]

We sample the posterior using Hamiltonian Monte Carlo in Stan \cite{Carpenter2017}. As a measure of evidence, we use the posterior probability that women are not disfavoured,
\begin{equation}
P(\alpha_j \geq 1) = 1 - \int_0^1 p(\alpha_j \mid r, \delta)\, d\alpha_j,
\end{equation}
and repeat the analysis for $0 \leq \delta \leq 20$ years.

\section*{Data availability}
The simulation code is available at \url{https://github.com/kaqmak/nobelgenderratio}. 


\end{document}